\documentclass[prd,12pt,aps,preprintnumbers,showpacs,tightenlines,floatfix,nofootinbib]{revtex4}
\usepackage{amssymb}

\def\Z{Z\!\!\!Z}
\def\GeV{~\mbox{GeV}}

\def\beq{\begin{equation}}
\def\eeq{\end{equation}}
\def\bea{\begin{eqnarray}}
\def\eea{\end{eqnarray}}
\def\bq{\begin{quote}}
\def\eq{\end{quote}}

\def\ba{\begin{array}}
\def\ea{\end{array}}

\newcommand{\aaa}{a}
\newcommand{\dd}{{\mathrm{d}}}

\begin{document}

\title{Towards understanding structure of the monopole clusters}

\author{M.~N.~Chernodub}
\affiliation{Institute of Theoretical and  Experimental Physics,
B. Cheremushkinskaja 25, Moscow, 117259, Russia}
\affiliation{Institute for Theoretical Physics, Kanazawa
University, Kanazawa 920-1192, Japan}
\author{V.~I.~Zakharov}
\affiliation{Max-Planck Institut f\"ur Physik, F\"ohringer Ring 6,
80805 M\"unchen, Germany}

\preprint{ITEP-LAT-2002-21}
\preprint{KANAZAWA-02-31}

\begin{abstract}
We consider geometrical characteristics of monopole clusters
of the lattice $SU(2)$ gluodynamics. We argue that the polymer
approach to the field theory is an adequate means to describe the
monopole clusters. Both finite-size and the infinite, or
percolating clusters are considered. We find out that the
percolation theory allows to reproduce the observed distribution
of the finite-size clusters in their length and radius.
Geometrical characteristics of the percolating cluster reflect, in
turn, the basic properties of the ground state of a system with a gap.
\end{abstract}


\date{\today}

\maketitle

\section{Introduction}
\label{one}

Explanation of the confinement in terms of the
monopole condensation was proposed as
early as around the year 1974~\cite{classics}. Moreover,
the idea is strongly supported by the lattice data~\cite{reviews}.
Nevertheless, understanding the lattice data in terms of the
continuum theory still represents a challenge.
Indeed, generically
one usually thinks in terms
of a Higgs-type model:
\begin{equation}\label{effective}
S_{eff}~=~\int d^4x\big({1\over 4}F_{\mu\nu}^2+|D_{\mu}\phi|^2+V(|\phi|^2)\big)
\end{equation}
where $\phi$ is a scalar field with a non-zero magnetic charge,
$F_{\mu\nu}$ is the field strength tensor constructed on the  dual-gluon
field $B_{\mu}$, $D_{\mu}$ is the covariant derivative
with respect to the dual gluon. Finally,
$V(|\phi|^2)$ is the potential energy ensuring that
$\langle\phi\rangle\neq 0$ in the vacuum.
There exist detailed fits to the
numerical lattice data within such a framework~\cite{action}.
However, relation of the ``effective''
fields $\phi,B_{\mu}$ to the fundamental QCD fields
remains unclear.

One of the main problems is the lack of understanding
of the nature of the monopole field configurations.
In particular, there is seemingly no reason  to think about
the monopoles as quasi-classical objects, for review see~\cite{alive}.
Under the circumstances, it is natural to exploit the lattice measurements
to accumulate first data on, so to say, phenomenology of the monopoles.
In fact, definition of the monopoles in the non-Abelian case is not unique
and we will concentrate on the monopoles defined within the Maximal
Abelian Projection (MAP) in SU(2) lattice gauge model.

A gauge-variant definition of the monopoles would not be of much interest by itself.
The basic idea is that there are gauge-invariant objects behind which are
detected through the projection.
This, gauge-invariant facet is manifested in gauge invariant properties
of the MAP monopoles.
In particular, (see~\cite{Bornyakov02} and references therein)
the three dimensional monopole density $\rho_{mon}$
does not depend on the lattice spacing and
is given in the physical units:
\begin{equation}
\label{scaling}
\rho_{mon}~=~ 0.65(2) ~\sigma^{3/2}_{SU(2)}~\approx~\mbox{const.}\,,
\end{equation}
where $\sigma_{SU(2)}$ is the string tension.

An important  remark is in order here. While discussing the
monopole density one should distinguish between finite-size
clusters and the percolating cluster~\cite{pochinsky,hart}. There
is a spectrum of the finite-size clusters, as a function of their
length, while the percolating cluster is in a single copy. In
other words, the percolating cluster fills in the whole of the
lattice and its length is proportional to the volume of the
lattice $V_4$: \beq L_{perc}~=~\rho_{mon}\cdot V_4 \eeq The
observation (\ref{scaling}) refers only to the percolating
cluster.

Also, upon identification of the monopoles in the Abelian
projection, one can measure the {\it non-Abelian} action associated with
these monopoles. The results~\cite{bornyakov01} turned in fact
astonishing and can be explained only as a fine-tuning~\cite{vz}.
To explain the meaning of the fine tuning, let us remind the reader
that the probability of finding any field configuration is
a product of action- and entropy- factors. Then, it turns out that for
the monopoles
at the presently available lattices
both factors diverge in the ultraviolet but cancel each other
to a great extent:
\bea
\label{tune}
P_{mon} = \exp(-S_{mon})\times (\mbox{entropy})~\sim \exp(-c_1\cdot L/a)\cdot
\exp(+c_2 \cdot L/a)\,,\\
{|c_1-c_2|\over a}~\sim ~\Lambda_{QCD}\,,
\eea
where $P_{mon}$ is the probability to find the monopole trajectory
of the length $L$, $a$ is the lattice spacing
and $c_{1,2}$ could in principle depend on the $a$ as well but are treated
as constants to describe the data on the available lattices
(with smallest lattice spacing $a \sim 0.06~\mbox{fm}$ from Ref.~\cite{bornyakov01}).

The observation (\ref{tune})
implies also that the monopoles look point-like on the
presently available lattices.
Indeed, let us normalize the phenomenological
monopole action for the lattice monopoles to the case
of point-like monopoles. For the Dirac monopole
the action would be:
\beq\label{actions}
S~=~L\cdot{1\over 8\pi}\int\limits_a^{\infty}
{\bf B}^2d^3r~=~\mbox{const.}\,{L\over ae^2}\,,
\eeq
where $e$ is the electric charge and
${\bf B}$ is the magnetic field of the monopole,
$B^2\sim 1/e^2r^4$.

Thus, at presently available lattices the MAP monopoles are
structure-less, or point-like objects. This observation
immediately triggers further questions since usually one considers
generic non-perturbative fluctuations as
large-scale. We would not discuss such questions
now  but simply stick to the phenomenological observation that at
the presently available lattices monopoles can be treated as
geometrical objects, with no size. One can advocate then~\cite{vz}
the use of the polymer approach to the field theory,
see, e. g., Ref.~\cite{parisi}.

Our basic idea here is to pursue a geometrical language to
describe monopoles, without any explicit use of Lagrangians
like (\ref{effective}).

In Sect.~\ref{two} we consider the relation between the
percolation theory and the monopole physics. In Sect.~\ref{three} we use the percolation
theory to evaluate the spectrum of the finite-size monopole clusters.
In Sect.~\ref{four} the same finite-size clusters are described in terms
of the simplest vacuum loop in the polymer representation.
In Sect.\ref{five} we comment on the properties
of the percolating cluster as representing
the ground state of the system. In Sect.\ref{six}
we discuss measurements which could clarify further
the structure of the effective theory describing the monopoles.
In Sect.\ref{seven} we present conclusions.

It is worth mentioning that in a few cases we
reproduce in some detail argumentation well known in other fields,
in particular in the condense matter.
And in this sense our presentation is not
original in these cases.
However, we believe that the justification of the use and application
of the condense-matter techniques to non-perturbative fluctuations
of the Yang-Mills fields is new.

\section{Monopole clusters and percolation}
\label{two}

In this section we will outline interconnections between
properties of monopole clusters, percolation theory~\cite{percolation} and field theory.
The aim is to motivate our basic assumptions (see subsection~\ref{two:six})
and to apply later the percolation theory to  the monopole clusters.

\subsection{Monopole condensation in compact $U(1)$}
\label{two:one}

A two-line theory of the monopole condensation was presented
first in Ref.~\cite{Polyakov75} for the case of the compact (or
lattice) $U(1)$ theory. We will review the derivation here and dwell
on its connection with percolation.

The role of the compactness of the $U(1)$ lattice
theory is to ensure that
the Dirac string does not cost any energy
(for a review see, {\it e.g.}, Ref.~\cite{alive}).
That is why in Eq.~(\ref{actions}) we take into account only the energy
of the radial field.
Nevertheless, the monopoles are
infinitely heavy in the limit $a\to 0$
and, at first sight, this precludes condensation since the
probability to find a monopole trajectory of the length $L$ is
suppressed as
\begin{equation}
\label{action}
\exp(-S)~=~\exp\left(-{c\over e^2}\cdot {L\over a}\right)\,.
\end{equation}
Note that the constant $c$ depends on details of the lattice
regularization but can be found explicitly in any particular case.

What makes the condensation feasible,
is an exponentially large enhancement factor due to
the entropy. Namely, a trajectory of length $L$ for a point-like
monopole can be
realized on a cubic lattice in $N_L=7^{L/a}$ various ways.
To evaluate the $N_L$
let us notice that
the monopole occupies the center of a cube and at each step
the trajectory can be continued to an adjacent cube.
In four dimensions there are 8 adjacent cubes.
However,
one of them is to be excluded from the counting
since the monopole world-line is
non-backtracking\footnote{If a piece of the trajectory is
covered in the both directions it is not observed on the
lattice. Physically, this cancellation corresponds to the cancellation
between monopole and anti-monopole.}. Thus the entropy factor is:
\begin{equation}\label{entropy}
N_L~=~\exp\left(\ln 7 \cdot {L\over a}\right)\,,
\end{equation}
and it cancels the suppression due to the action
(\ref{action}) if the coupling $e^2$ satisfies the condition of criticality:
\begin{equation}
\label{critical}
e^2_{crit}~=~c/\ln7~\approx~1\,
\end{equation}
where we quote the numerical value of $e^2_{crit}$ for the Wilson action
and cubic lattice. At $e^2_{crit}$  any trajectory length
$L$ is allowed and the monopoles condense.
This simple framework is verified within  about one percent accuracy
as far as the prediction of $e^2_{crit}$ is concerned~\cite{BanksMyersonKogut,shiba}.

One can say that the coupling $e^2$ of
the compact $U(1)$ is to be fine-tuned to
trigger the phase transition.

\subsection{Relation to percolation}
\label{two:two}

In fact the derivation
of the preceding subsection can be viewed
as an application of the percolation theory~\cite{percolation}. Moreover, one thinks in terms of
the simplest percolation possible, that is, uncorrelated percolation.

Indeed, monopoles are observed as  trajectories on
the links of the dual lattice~\cite{reviews}.
Postulate that the probability to occupy a link is  given by:
\beq\label{p}
p~=~\exp(-c/e^2)~,
\eeq
compare Eq. (\ref{action}) at $L=a$.
The probability that (uncorrelated) links form a trajectory of
length $L$ is then given by Eq. (\ref{action}).

Formation of an infinitely long, or percolating cluster
at a critical value $p=p_{crit}$  is a common
feature of percolating systems. In our case,
\beq\label{pc}
p_{crit}~=~1/7~,
\eeq
see Eq. (\ref{critical}).

It might worth mentioning that in text-books one would rather find
$p_{crit}=1/8$ since the non-backtracking feature of the monopole trajectory
is specific for charged particles.
From the theoretical point of view the non-backtracking
is a manifestation of the monopole charge conservation.
Another consequence of this conservation law is the closeness of the trajectories
which has not been taken into account so far. As we shall see in Sect.~\ref{four}
the closeness of the trajectories brings in
only a pre-exponential factor
and can be ignored at the moment for this reason.

\subsection{Relation to field theory}
\label{two:three}

The derivation in Section~\ref{two:one} implies that the monopole condensation
occurs when the monopole action is ultraviolet divergent.
On the other hand, the onset of the condensation in
the standard field theoretical language corresponds to
zero mass of the magnetically charged field $\phi$.
It is important to realize that this apparent
 mismatch between the two languages
is not specific for the monopoles at all. Actually, there is
a general kinematic relation (see, e.g., Ref.~\cite{samuel79,ambjorn}) between
the physical mass of a scalar field $m^2_{prop}$ and the mass
$M$ defined in terms of the (Euclidean) action $S$, $M\equiv S/L$:
\begin{equation}\label{massrenormalization}
m^2_{prop}\cdot a~\approx~C^2_m\cdot \big(M(a)-{\ln 7\over
a}\big)\,,\quad C_m^2 = 8\,.
\end{equation}
It is $m^2_{prop}$ that enters the propagator of a scalar particle
$D(p^2)$,
\beq\label{m2}
D_{sc}(p^2)~\sim{1\over p^2+m^2_{prop}}\,,
\eeq
not $M^2(a)$. The factor  $\ln 7$ in
Eq.~(\ref{massrenormalization}) is specific for the ``monopole walk''
(see the preceding subsection). The Eq.~(\ref{massrenormalization})
establishes a connection between uncorrelated percolation and propagator
of a free particle (in Euclidean space),
$$C^2_m \, m^2_{prop}a^2~\approx~\ln(p_{crit}/p),~~~p_{crit}>p\,,$$
where we neglect, as usual, higher corrections in $a$.

It is also worth mentioning that the overall normalization of the
propagator of a free particle in the polymer approach reads in the
leading (in the lattice units $a$) order as follows:
\beq\label{path}
\sum_{paths}\exp\Bigl\{-M(a)L_{path}\Bigr\} =
C_m^2 \, a^2 \, D_{sc}(x_i,x_f)\,,
\eeq
where $D_{scalar}(x,x^{'})$  is the standard field-theoretical propagator
of a scalar particle with mass
(\ref{massrenormalization}) in the Euclidean space-time,
$L_{path}$ is the length of a particular path connecting
the initial and final points $x_{i,f}$ .
Note that the action factor for a given path, $\exp(-M\cdot L_{path})$,
is similar to the action of a polymer of length $L$, with
the chemical potential replaced by the mass $M(a)$

Eq.~(\ref{massrenormalization}) demonstrates that the propagating
mass can be kept finite in the limit $a\to 0$ only at price of
fine tuning. Moreover, Eq.~(\ref{massrenormalization}) looks
exactly the same as the standard radiative correction to the mass
of a scalar particle. In more detail, the first term in the
right-hand side is actually the magnetic field energy. The $\ln 7$
term, on the other hand, plays the role of a counter-term and this
counter-term is calculable in our case!

\subsection{Percolation and the quadratic divergence in field theory}
\label{two:four}

Now we come to explain an important difference between
phase transitions in case of uncorrelated percolation and in case of the
monopoles. This difference might come as a kind of surprise since the
transition point can be determined within the percolation theory
with a good accuracy (see the discussion of the $U(1)$ case above).

The difference concerns properties of a percolating
system and of the  monopole vacuum
in case of $p>p_{cr}$ and in the confining phase, respectively,
If $p>p_{cr}$ the product of
probability and entropy factors  reduces to
$\exp(+|\varepsilon|\cdot L)$, where $\varepsilon$ can be
arbitrarily small. Then, it is clear that the longest possible
trajectory wins and the cluster fills in a finite fraction
of the whole lattice.
In the percolation-theory language
the statement is that the fractal dimension at $p>p_{crit}$ coincides
with the space dimension~$D$:
\beq\label{fractals}
(D_{fr})_{perc}~=~D~,~ p>p_{crit}~,
\eeq
see, {\it e.g.}, Ref.~\cite{aizenman} and references therein.
Note that in the language of field theory
$p>p_{crit}$ corresponds to a tachyonic
mass, $m^2_{prop} < 0$, see Eq.~(\ref{massrenormalization}).

In case of monopoles, Eq (\ref{fractals})
would correspond to
\beq\label{divergence}
(\rho_{mon})_{perc}~~\sim ~~a^{-3}~\,.
\eeq
Note that in the field theoretical language Eq.~(\ref{divergence})
would be interpreted as a power-like divergence well known in perturbation
theory of charged scalar particles. Indeed, already on dimensional
considerations one would conclude that:
\beq \label{natural}
~\langle|\phi|^2\rangle_{perc}~\sim~a(\rho_{mon})_{perc}~~\sim ~~a^{-2}~,
\eeq
(for a more careful derivation see~Ref.~\cite{vz}).

The behaviour (\ref{divergence}) is in a sharp
contrast with experimental data, see (\ref{scaling})
and this is what we mean by the difference in the behaviour
of a percolating system at $p>p_{cr}$ and of the monopoles in
the confining phase. In other words, the fine tuning exhibited by
the monopoles in the confining phase is a specific feature of the
vacuum state of the lattice gauge theory.

\subsection{Fine tuning and Coulomb-like interaction}
\label{two:five}

Let us emphasize again that in the present note we treat the fine
tuning of the monopoles as a pure observation and look for its
implications. We are not in position at all to track its origin
back to the non-Abelian Lagrangian which after all determines the
results of all the lattice measurements. One of very few
theoretical points which we can nevertheless add is that a
long-range, Coulomb-like force is a necessary ingredient of a
fine-tuned theory. Of course, in case of color by long
distances we mean  $r\gg a$ but $r \le \Lambda_{QCD}^{-1}$.

To substantiate the point, let us consider the
confining phase of the compact $U(1)$.
The relative simplicity of this case is that the
dynamics is explicitly known and, moreover, determined by the
value of the electric charge $e$ which can be tuned arbitrarily ``by
hand'' (while in the non-Abelian case the coupling runs and is not
under control in this sense).

Thus, one can introduce two scales in the confining phase of the compact
$U(1)$ in the following way:
\beq
R_{UV}~=~a~,~~R_{IR}~=~{a\over \epsilon}~~,~~~\epsilon~\equiv~
(e^2-e^2_{crit})~>0~;~~
\epsilon~~\ll ~1~\,.
\eeq
At first sight, we are in the same situation as in case of
uncorrelated percolation since
we introduced a tachyonic mass,
$$e^{-S}~\sim~e^{-\mu\cdot L}~,~~~\mu~~\sim~-{\epsilon\over a}~, $$
and nothing can prevent development of a quadratic divergence,
see discussion above.

On the other hand it is known from the lattice measurements
(see, {\it e.g.}, Ref.~\cite{u1} and references therein) that the IR scale
does coexist with the UV one. And, indeed, there is the loophole in our
logic that introducing the tachyonic mass would immediately result
in a  quadratic divergence. This would be true if the interaction
were local. However, imagine that there are other monopoles
at a distance of order $R_{IR}$. Then these monopoles can modify the
Coulomb field of the monopole considered by order unit at
$r\sim R_{IR}$. This, in turn,
can be interpreted as
a change in the mass (due to interactions):
\beq\label{delta}
|\delta M(a)|~\sim~{\epsilon \over a}.
\eeq
This change can compensate for the ``short-distance''
tachyonic mass so that the particles are kept away from each other
at a distance of order $R_{IR}$. Which also means that
there is no power-like UV divergences (see discussion above).
This double-face interpretation of (\ref{delta}) as IR and UV effects is
a unique feature of the Coulomb-like interaction.

Note that we are not really giving a proof that the introduction
of $R_{IR}$ is self-consistent. We just argue that {\it without}
long-range forces the fine tuning would be not possible at
all. In
case of non-Abelian theories there are many more open questions
since the monopole action is not bounded from below (see,
{\it e.g.}, Ref.~\cite{alive}) and the fact that the action is tuned to the
entropy is even more difficult to explain theoretically.

\subsection{Fine tuning and $\lambda\phi^4$ models}
\label{two:six}

It is worth to emphasize that the fine tuning which is observed
for the lattice monopoles is of the same generic type considered
so mystifying in case of the Standard Model. There is, however,
a peculiarity in our case. Indeed, if we compare the ``natural'' estimate
for $\rho_{mon}$  (\ref{divergence}) with the data (\ref{scaling})
we see that there are three powers of $a^{-1}$ that are
``tuned away'', not two powers as in the standard field-theoretical
language. And, indeed, the data (\ref{scaling}) imply in the ``naive''
$a\to 0$ limit~\cite{vz}:
\bea\label{calm}
\lim_{a\to 0}\langle|\phi|^2 \rangle ~\sim~\rho_{mon}\cdot a~\to0\,,\nonumber\\
\lim_{a\to 0} m_{prop}^2~\sim \lim_{a\to 0}\frac{1}{a}{\big(M(a)-\ln7/a\big)}~\sim
~{\Lambda_{QCD}\over a}~\to~\infty\label{unusual}\\
\lim_{a\to 0}m_{gluon}^2~\sim ~g^2 \langle |\phi|^2 \rangle ~\to ~0~\,, \nonumber
\eea
where $m_{gluon}$ is the (dual) gluon mass which
arises in theories of the type (\ref{effective}).

Eqs.~(\ref{unusual}) are of course not what one would expect
for the standard $\lambda\phi^4$ theory with spontaneous symmetry breaking.
It is worth emphasizing, however, that the limit  $a\to 0$ in Eqs.~(\ref{unusual})
should be understood with some reservations. What we mean actually
in Eqs.~(\ref{unusual}) by $a\to  0$ limit
is ``the lattice spacing
$a$ as small as possible within availability on present lattices''. The behaviour
(\ref{unusual})
can actually change in the academic limit $a\to 0$
(i.e. at lattice spacings much smaller than those presently available).
Subsequent considerations
in the present paper would actually suggest such a possibility (see Sect.~\ref{five}).

Although Eqs.~(\ref{calm}) may not survive at smaller $a$, they do answer questions
why new point-like fluctuations (implied by the fine-tuning)
do not disturb the standard $\beta$-function of the $SU(2)$ gluodynamics
at existing lattices.
Indeed, according to (\ref{calm}) all the scalar degrees of freedom are actually
removed from the physical spectrum  if $a\to 0$.

The properties (\ref{calm}) imply also that the potential energy
$V(|\phi|^2)$ scales in the limit $a\to$ and our effective theory
can well be a useful approximation to study the vacuum properties.
In particular, the monopole confining mechanism survives in the
limit~\cite{vz} $a\to 0$.

\subsection{Formulating the main hypothesis}

After all these preliminary discussions we are set to formulate our
main hypothesis.
Namely, we will assume that we can consider the point
\beq\label{pcr}
p~=~p_{crit}
\eeq
as adequately describing the physics in the {\it confining phase}
of the non-Abelian gauge theory. The justification for
this hypothesis is the fine tuning observed on the
lattice and discussed in length above.

There are important reservations to be made.
Namely, the fine tuning does
imply that the physics is so to say ``frozen'' at $p=p_{cr}$ as
far as the UV scale is concerned. However, as far as the
dependence on the scale $\Lambda_{QCD}$ is concerned it could be
different than at the point (\ref{pcr}). Moreover, we can give a
more quantitative meaning to the scale ``$\Lambda_{QCD}$'' in the
case considered. The point is that the percolating cluster has
self-crossings and the length of the trajectory between these self-crossings
can be considered as the monopole free-path length.
Direct measurements show~\cite{boyko} that this
length (measured along the trajectory) scales:
\beq\label{travel}
L_{free}~\approx~1.6~fm\,.
\eeq
Thus, we can apply our hypothesis as long as
\beq\label{limits}
L_{free}~\gg~L~\gg~a\,.
\eeq

In the next section we exploit the percolation theory
to describe the structure of the finite-size monopole
clusters satisfying (\ref{limits}).

\section{Finite-size clusters and percolation}
\label{three}

\subsection{Data and percolation picture}

Detailed data on the structure of the monopole clusters were
obtained in~\cite{hart}. As was mentioned above, there is a
 single percolating cluster, whose length grows with the
lattice volume, and finite-size clusters. In this section we will
concentrate on the finite-size clusters
satisfying the condition (\ref{limits}).
These clusters are characterized, first of all, by their length.
It was found that the length spectrum is described by a power law:
\beq\label{spectrum}
N(L)~=~{c_4\over L^{\tau}}\cdot V_4~,
\eeq
where
\beq\label{gamma}
\tau~\approx~3
\eeq
for all lattice spacings  and sizes tested and the coefficient $c_4$
depends only weakly on $\beta$.
For our purposes we can neglect this dependence.

Another important characteristics of the clusters is their radius
$R_L$ as function of the length $L$. By the radius one understands
the average distance between two cluster links:
\beq\label{radii}
R_L^2~=~{a\over L}\sum_{i=1}^{L/a}({\bf x}_i-{\bf \bar{x}})^2~=
{a^2\over 2L^2}\sum_{i,j=1}^{l/a}|{\bf x}_i-{\bf x}_j|^2~~,
\eeq
where ${\bf x}_{i,j}$  are coordinates of the links
and $\bar{\bf x}~=~(1/L)\sum {\bf x}_i$.
The measurements indicate:
\beq\label{radius}
R_L~\sim~\mbox{const}_1 + \mbox{const}_2\sqrt{L}\,.
\eeq

Thus, our problem is to clarify whether the data (\ref{gamma})
and (\ref{radius})
can be understood within the percolation theory (and our main hypothesis, see
subsection~\ref{two:six}). It is encouraging to observe
that even without any dynamical
input we can conclude that the lattice data on the finite-size monopole
clusters reveal a picture typical for percolating systems.
Indeed, a generic form of the spectrum, for $p< p_{crit}$ is
\beq\label{assumption}
N(L)~\sim~{\exp(-\mu\cdot L)\over L^{\tau}}\cdot V_4\,,
\eeq
where $\tau$ is the so called Fischer exponent and $\mu$ vanishes
at the critical point, $p=p_{crit}$. The latter is easy to understand since at the
point of the phase transition
the correlation length is infinite and
there is no dimensional parameter left.
Moreover, a power law like (\ref{radius}) is common to the percolating systems,
\beq
R_L~\sim ~L^{1/D_{fr}}\,,
\eeq
and $D_{fr}$ is called the fractal dimension.
For the monopoles we apparently have from the lattice measurements:
\beq\label{fractal}
(D_{fr})_{mon}
~\approx~2\,.
\eeq

Thus, the data on the finite-size monopole clusters exhibit a typical
percolation picture and the next question is whether it is possible
to evaluate the exponents (\ref{gamma}) and (\ref{fractal}).

\subsection{Hyperscaling relation}

Similarity between percolation and properties of dynamical systems
undergoing the phase transition is well known~\cite{percolation}.
However,
direct evaluation of the critical exponents requires, generally speaking,
a particular dynamical input on the system considered.
Mostly, knowledge on the excitation spectrum is required.

However, there is a  general relation between the critical
exponents which we are going to apply first.
Note that our considerations here are not specific for the monopoles and
similar, {\it e.g.}, to the analysis of Ref.~\cite{sch}.
Still, for the sake of orientation let us mention a few points
important for the application of the percolation theory to
the monopoles clusters.

First, one derives all relations for $p \le p_{cr}$, without considering
$p>p_{cr}$. As we discussed in length in Sect.~\ref{two}, uncorrelated percolation can
be indeed relevant to the monopole physics only at $p\le p_{cr}$.
Second, all the relations of the percolation theory follow more or less
directly from the assumption (\ref{assumption}) and it is worth emphasizing
that the spectrum (\ref{assumption}) is very natural physics-wise. Indeed,
the presence of the exponential
factor at $p-p_{cr}<0$ is obvious from considerations of subsection~\ref{two:two}.
Moreover at the point $p=p_{cr}$ there is no scale left and the power like
dependence of the spectrum on $L$ is the only dependence allowed. In this
way one comes to include the factor $L^{-\tau}$ as well.

Starting from the spectrum (\ref{assumption}) it is quite
straightforward, see, {\it e.g.}, Ref.~\cite{sch}, to derive
the following well-known relation:
\beq\label{hyperscaling}
\tau~=~{D\over D_{fr}}+1~~,
\eeq
where $d$ is the dimension of space-time and $D_{fr}$ is in fact $D_{fr}(p=p_{cr})$.

In our case,
\beq\label{main}
\tau~\approx~3~,~~D~\equiv~4,~~D_{fr}~\approx~2\,.
\eeq
Thus, Eq. (\ref{hyperscaling}) is satisfied within
the error bars of the lattice measurements
and this observation is one of our main results.
In view of its importance, we will later rederive  (\ref{main})
and some generalizations of it in the language of field theory.

\subsection{Fractal dimension}

As is mentioned above, the relation between the length and radius
of the cluster is determined by the fractal dimension,
see Eq.~(\ref{radius}). The fractal dimension, in turn,
is determined by the kind of the walk.

In fact we are dealing with the {\it monopole walk} which, to the
best of our knowledge, has not been studied in detail in the
literature. However, the
characteristics of the monopole walk are so to say flanked by the
characteristics of the well known {\it random} and {\it
self-avoiding} walks. Indeed, the monopole walk chooses freely one
of 7 directions available at each step. This is in common with the
random walk. On the other hand, choosing the eighth direction
would result in an immediate self-crossing and this is forbidden.
The latter feature is in common with the self-avoiding walk. However,
in contradistinction from the self-avoiding walk, self-crossings
are allowed for the monopole trajectories at later stages.

The observation central for this section is that in $D=4$ the
fractal dimension $D_{fr}=2$ both for the random and self-avoiding
walks. Therefore we can predict $D_{fr}=2$ for the monopole walk
as well.

In more detail, for the random walk one has in any number of space
dimensions \footnote{ This relation, in connection with the
monopole clusters, is in fact mentioned in Ref.~\cite{hart}.}:
\beq
(D_{fr})_{random}~=~2\,.
\eeq
For the self-avoiding walk
one has (the so called Flory's fractal
dimension):
\beq
(D_{fr})_{self-avoiding}~=~{D+2\over 3}
\eeq
which is valid for $D\le 4$ and gives exactly the same $D_{fr}=2$ at $D=4$.

Thus, the experimental value
$(D_{fr})_{mon}=2$ appears to be well understood theoretically.

\section{Seeing  free monopoles at short distances}
\label{four}

From our discussion of the relation between the uncorrelated percolation
and free-field theory (see Sect.~\ref{two:three}) it is clear that the results of the
preceding section can be derived in terms of Feynman graphs as well
\footnote{We are grateful to D. Diakonov for a discussion
on this point and providing us with a reference to \cite{sasha}.}.
Moreover, since we are considering relatively short trajectories,
see (\ref{limits}), the effect of the mass can be neglected.
In this section we will reproduce the results above by
considering the
simplest Feynman graph relevant, that is a single vacuum loop.
 This approach  allows us to
derive also some generalizations, like
length-of-the-trajectory spectrum at finite temperature.

\subsection{Polymer representation for a massless particle propagator}
\label{four:one}

In this subsection we will evaluate
the Fischer exponent $\tau$ (see (\ref{hyperscaling}))
starting from the observation that
the fluctuations of the field $\phi$ are massless on the scale $a$.
To this end we  reproduce first the basic points of
the polymer approach to the field theory of a free scalar
particle, see, {\it e.g.}, Refs.~\cite{stone78,samuel79,sasha,ambjorn}.

The partition function for a closed polymer is:
\bea
Z = \int \dd^4 \, x \, \sum^\infty_{N=1} \frac{1}{N} \, e^{ - M\cdot N}
\, Z_N(x,x)\,,
\label{Z}
\eea
where $M$ is the chemical
potential and $Z_N(x_0,x_f)$ is the partition function of a polymer
broken into $N$ segments:
\beq
Z_N(x_0,x_f) = \Bigl[\prod\limits^{N-1}_{i=1} \int \dd^4 x_i\Bigr] \,
\prod^{N}_{i=1} \Biggl[\frac{\delta(|x_i - x_{i-1}|-a)}{2\pi^2
a^3}\Biggr]\,
\label{ZN}
\eeq
This partition function (\ref{Z}) contains a summation over all atoms of
the polymer weighted by the Boltzmann factors.
The $\delta$--functions in (\ref{ZN}) ensure that each bond in
the polymer has length $a$. The starting point of the polymer
is $x_0$ and the ending point is $x_f \equiv x_N$.

Note that there is a pre-exponential factor $1/N$ in Eq.~(\ref{Z}).
This is due to considering closed trajectories.
Indeed, the factor is introduced
to compensate for the N-multiple counting of the same closed trajectory
in the partition function (\ref{Z})
since any atom on this trajectory can be considered as the initial and
final point. As we shall see later, our final result crucially depends
on the pre-exponential factors.

The crucial step to relate (\ref{ZN}) to a free particle path integral
is the so called coarse--graining. Namely, the $N$--sized polymer
is divided into $m$ units by $n$ atoms ($N = mn$), and the limit is considered
when both $m$ and $n$ are large while $a$ and $\sqrt{n} a$ are
small. We get,
\beq
\label{constraint}
\prod\limits^{(\nu+1)n-1}_{i=\nu n} {1\over
2\pi^2\aaa^3}\delta(|x_i-x_{i+1}|-\aaa) \to
{\Bigl(\frac{2}{\pi n a^2}\Bigr)}^2
\,\exp\Bigl\{ - \frac{2}{n \, a^2} {(x_{(\nu + 1) n} - x_{\nu n
})}^2 \Bigr\} \,,
\eeq
where the index $i$, $i=\nu n \cdots (\nu+1)n-1$, labels the atoms in
$\nu^{\mathrm{th}}$ unit. The polymer partition function
becomes~\cite{stone78}:
\bea
\label{polymer}
Z_N(x_0,x_f) & = & {\mathrm{const}} \cdot
\left[
\prod^{m-1}_{\nu =1} \dd^4 x \right] \Biggl[
{\Bigl(\frac{2}{\pi n a^2}\Bigr)}^{2 m} \exp\Bigl\{ -2
\sum_{\nu=1}^m{(x_{\nu}-x_{\nu-1})^2\over n\aaa^2}\Bigr\}\Biggr]
\nonumber \\
& & \cdot \exp\Bigl\{ - \sum_{\nu=1}^m n\cdot a\,\mu  \Bigr\}\,.
\eea
The $x_i$'s have been re-labeled so that $x_{\nu}$ is the average
value of $x$ in at the coarser cell. Note also that at this stage
there appears the
chemical potential $\mu$ related to the
original parameter $M$ through
\begin{equation}
\label{mu}
\mu = M-\ln 7/a\,,
\end{equation}
as is discussed above.

Using the variables:
\beq
\label{related}
s={1\over 8}\,n\aaa^2\nu\,,\quad l={1\over 8}\aaa^2\,N\,, \quad m_{prop}^2=
{8\mu \over \aaa}\,,
\eeq
one can rewrite the partition function (\ref{Z}) as
\beq
\label{m0}
Z = {\mathrm{const}}\cdot \!\int\limits_0^{\infty}
\frac{\dd l}{l} \! \!\!\int\limits_{x(0) = x(l)=x}\!\!\!\!\!\!\!\!\!
D x\, \exp\Biggl\{-\int\limits_0^l \Bigr[{1\over 4}\dot{x}^2_\mu(s)
+ m_{prop}^2
\Bigr] \, \dd s \Biggr\}\,.
\eeq
Note that the mass renormalization in Eq.~(\ref{related}) is
consistent with Eqs.(\ref{massrenormalization},\ref{mu}).

 After these preliminary steps we can readily derive the distribution
in the length of the trajectories in the massless case.
To this end, let us rescale $x _{\mu}$ and $s$ in such a way,
 that
there is no $l$ dependence left in the action if $m_{prop}=0$:
\beq\label{rescaling}
L~=l/a,~~\tilde{s}~=~s/l,~\tilde{x}_{\mu}~=~x_{\mu}/\sqrt{l}~\,.
\eeq
Then, indeed,
\beq\label{subtle}
Z = {\mathrm{const}}\cdot \!\int\limits_0^{\infty}
\frac{\dd L}{L}\cdot~ I~,
\eeq
where
\beq
I~\equiv~\int\limits_{\tilde{x}(0) = \tilde{x}(l)=\tilde{x}}
D \tilde{x}\, \exp\Biggr\{- {1\over 4} \int\limits_0^1\dot{\tilde{x}}^2_\mu(\tilde{s})
 \, \dd \tilde{s} \Biggr\}\,.
\eeq
At first sight, Eq.~(\ref{subtle}) implies that we have a $dL/L$ spectrum since there is
no $L$ dependence left otherwise. However,
the actual spectrum refers to the number of loops in
a unit volume (see (\ref{spectrum}) and the discussion of it).
Thus, the $dL/L$ spectrum refers to the volume in the $\tilde{x}$ units.
Since the $\tilde{x}^D~\sim~x^D L^{-D/2}$
we have in the physical-volume units:
\beq\label{nice}
N(L)dL~=c_D~V_D ~{dL\over L^{1+D/2}}~,
\eeq
where $V_D$ is the volume in $D$-dimensional space and $c_D$ is a constant.
Although we are interested only in the $D=4$ case we kept $D$ as variable to
emphasize that we rederive in fact the hyperscaling relation
(\ref{hyperscaling}). Note that $D_{fr}=2$ is implicit in our derivation
and is encoded in fact in the transformation (\ref{constraint}).

\subsection{Coulomb-like interaction}

So far we considered approximation of free particles which corresponds to
the dominance of the monopole self-energy.
We expect, however, that the monopoles
interact also Coulomb-like.
Other, effective interactions are not ruled out either.
The Coulomb-like interaction can readily be included into the
action in the polymer representation. The corresponding extra piece
in the action is given by:
\beq\label{coulomb}
S_{Coulomb}~=~{g_M^2\over 2}\int_0^l\int_0^l ds_1 ds_2 \dot{x}_{1,\mu}D_{\mu\nu}
(x_1-x_2)\dot{x}_{2,\nu}
\eeq
where $g_M$ is the magnetic charge,
$D_{\mu\nu}(z)$ is the massless photon propagator, and the dot, as usual,
means differentiation with respect to the proper time.

The action (\ref{coulomb}) is manifestly invariant under the rescaling
(\ref{rescaling}):
\beq
S_{Coulomb}~=~{g_M^2\over 2}\int_0^1\int_0^1 d\tilde{s}_1
d\tilde{s}_2 \dot{\tilde{x}}_{1,\mu}D_{\mu\nu}
(\tilde{x}_1-\tilde{x}_2)\dot{\tilde{x}}_{2,\nu}~,
\eeq
and, therefore, the $1/L^3$ behaviour of the spectrum (\ref{nice})
should be still true for\footnote{The case $D=4$ is special
because the gauge coupling is dimensionless.} $D=4$ upon inclusion
of this interaction\footnote{We do not discuss here subtleness
which can arise from consideration of the coinciding points,
$x_1\to x_2$, in Eq.~(\ref{coulomb}).}.

\subsection{Finite temperatures, finite volume lattices}

Formalism of Section~\ref{four:one} can readily be generalized to the case of
a non-zero temperature. Indeed, we should rewrite now in terms of
a length-of-the-trajectory distribution the propagator of a
massless particle at finite temperature.

As usual,  finite temperature in Euclidean space corresponds to a
compactified fourth direction. Thus, we simply write the
$L$-distribution corresponding to an ensemble of non-interacting
particles with masses $m^2=(2\pi n)^2T^2$ in d=3 space--time:
\beq\label{temperature}
N(L)~=~c_3~
V_3~{1\over L^{5/2}}\sum_{n\in\Z}\exp\big(-(2\pi
nT)^2L\cdot a\big)\,,
\eeq
where $V_3 = V_4 T$ is the three dimensional volume of the
time--slice. Note that in the limit $T\to 0$ the sum over the exponentials
is proportional to $1/(T\sqrt{L\cdot a})$ and we come back to (\ref{nice})
upon proper normalization of the constant $c_3$,
$$c_3~=~2\,c_4\,\sqrt{\pi a}\,,$$
where $c_4$ enters Eq.~(\ref{spectrum}). However, in the opposite
limit, $T\to\infty$ we effectively get three dimensional theory
with the volume $V_3$ and the loop distribution~$N(L) \propto L^{-5/2}$.

A remark on the range of validity of (\ref{temperature}) is now in
order. The point is that in (\ref{temperature}) we disregard
chemical potential $\mu\neq 0$. Whether this is a valid
approximation depends on the numerical values of the parameters
involved, $\mu,T,a$. In particular, if we tend $a\to 0$ and keep
$\mu$ in physical units the approximation (\ref{temperature}) is
not valid.

Proceeding in a similar way we can derive also the spectrum on
a finite-size lattice of the volume $V_4 = \prod^4_{\mu=1} X_\mu$, where
$X_\mu$ is the length of the space in $\mu^{\mathrm{th}}$
direction. Assuming the periodic boundary condition we get:
\beq
N(L)~=~{c_0\over L} \prod^4_{\mu=1} \sum_{n_\mu \in \Z}
\exp\big\{-(2\pi n_\mu \slash X_\mu)^2 L\cdot a\big\}\,,
\eeq
where
$$c_0~=~16 \pi^2 \, c_4 a^2\,.$$

\section{Infrared cluster and properties of the ground state}
\label{five}

In this section we discuss geometrical properties of the
percolating cluster. It was suggested already in \cite{stone78,samuel79} that
the percolating cluster corresponds to the $\phi$-field condensate
in the classical
approximation \footnote{Literally, a model with a tachyonic mass was
introduced first in~\cite{stone78} and such a model
would result in $\rho_{mon}\sim a^{-3}$.}. However, it is only the phenomenon
of the fine tuning that makes the properties of the vacuum state non-trivial.
Indeed, we have now two coexisting scales, $a$ and
$\Lambda_{QCD}^{-1}$. Moreover, we will see that the properties of the
percolating cluster differ radically from the properties of the finite-size
clusters.

\subsection{Lattice data}

We have already mentioned that the density of the monopoles in
the percolating cluster scales, see (\ref{scaling}).
Recently, further measurements on the geometrical
elements of the percolating cluster have been performed~\cite{boyko}.
In more detail, the cluster consists of self-crossings and
segments connecting the crossings.
It was found that the average distance between
the crossings measured along the trajectory scales, see Eq.~(\ref{travel}).
Violations of the scaling in $L_{free}$ are negligible for all
the lattices tested.

One can also measure the average of the shortest, or  Euclidean
distance, $\langle d \rangle$, between the two crossings connected by
segments. In this case violations of the scaling are more
significant and, roughly, the data can be approximated as:
\beq\label{d}
\langle d \rangle \sqrt{\sigma}~\approx~
\big(0.65~+~[(a\sqrt{\sigma})-0.25]\big)~,
\eeq
where $a$ is the lattice spacing, $\sigma$ is the string tension
for the $SU(2)$ gluodynamics, and $0.15~<~a\sqrt{\sigma}~<0.35$.

Thus, we are confronted with the problem of interpreting the data
(\ref{scaling}), (\ref{travel}), (\ref{d}). The first step is to
reduce the set of data to a simple picture: normalized to a
free-particle case the measurements correspond to an infinite
mass~\cite{boyko}, $m^2_{prop}\sim a^{-2}$. Which is at first
sight a shocking observation defying everything that we said so
far. Postponing discussion of the physics till the next subsection
let us reiterate the reasons for such an interpretation of the
data.

For a free particle, one can measure its mass by comparing the
Euclidean distance between two points with the length along the
corresponding trajectory connecting the same points. Indeed, one
obtains the distance $L$ by differentiating the partition function
(\ref{Z}) with respect to the chemical potential $\mu$ defined
in Eq.~(\ref{mu}).
On the other hand, the dependence of a free particle propagator on
the chemical potential enters through the factor
$\exp(-m_{prop} \, d)$ where the $m_{prop}$ is as in
Eq.~(\ref{massrenormalization}).
Differentiating with respect to $\mu$
the propagator $D(m_{prop},d)$ of a free particle of mass
$m_{prop}$ we get therefore:
\beq\label{shortdistance}
\langle L_{free~particle}\rangle~= -{\partial\over \partial \mu}
\ln D(m_{prop}, d)~ \approx~{d_{free~particle}\over 2(m_{prop}(a)\cdot
a)}\,,
\eeq
where we neglected higher corrections in $a$ and $m_{prop}$ (generally
speaking, $m_{prop}$ also depends on $a$).

Two particular cases are worth mentioning. First, if $m^2_{phys}$ is in the
physical units, $m^2_{prop}\sim\Lambda_{QCD}^2$ then
we would have
$$\langle L \rangle~\sim~{\langle d \rangle\over a}$$
which is in blatant disagreement with the data (\ref{d}).
Moreover if we assume cancellation of only the
leading $1/a$ term in (\ref{massrenormalization})
we would expect:
$$\langle L \rangle ~\sim {\langle d \rangle \over \sqrt{a}}~,$$
which still cannot be reconciled with the data either.

Finally, if $m^2_{prop}\sim a^{-2}$ then
$$\langle L \rangle~\sim~\langle d \rangle$$
which does agree with the data (\ref{d}) for
$a$ available as far as we neglect the correction linear in $a$.

To summarize, we are coming to a paradox. Indeed, even if we
accepted $m^2_{prop}\sim a^{-2}$ to satisfy $\langle L \rangle
\sim \langle d \rangle$ this would not help since we would not be
able then to explain the persistence of the physical scale in
(\ref{travel}), (\ref{d}).

\subsection{Analogy to the M\"o\ss bauer effect}

Thus, we see that the finite-size clusters
and the percolating cluster exhibit
different patterns of the monopole kinematics.
In the former case we could neglect the monopole mass
while in the latter case it is not possible at all.
Looking for an analogy, we naturally
come to the M\"o\ss bauer effect. Indeed, one could say that the effect is
the difference in kinematics inherent to the decays of a
free atom and of an atom
belonging to a  lattice in its ground state.
Formally, the kinematics of the decay of the atom belonging to
the lattice looks as if the atom had an infinite mass.
Thus, the analogy we are going to pursue is between the percolating
monopole cluster and atoms in the ground state of a lattice of atoms.
The reason for the analogy is that in the both cases we are dealing
with ground states of systems with a gap.

 Let us first remind the reader the basic features of the M\"o\ss bauer
phenomenon (for a review see, {\it e.g.}, Ref.~\cite{lipkin}).
One considers decay of an atom belonging to a lattice of atoms.
Then there exists a probability $P_0$ that the lattice is not excited
at all, i.e. no phonons are emitted. Then the recoil momentum
is transferred to the whole of the lattice and, clearly, this
corresponds to an infinite effective mass of the decaying atom.

In the language of the quantum mechanics, the probability
of the M\"o\ss bauer transition
is determined by the following matrix element:
\beq
P_0~=~|\langle i|\exp(i{\bf p_{\gamma}}\cdot {\bf X}_{atom})|i\rangle|^2
\eeq
where $|i \rangle$ is the ground state of the lattice,
${\bf X}_{atom}$ is the position of the decaying atom and
${\bf p}_{\gamma}$ its recoil momentum.
Then, there are two crucial quantities which determine the estimate
of $P_0$, namely the ``naive'' recoil energy $R$ and the energy
of the gap. For non relativistic kinematics,
$$R~=~{\bf p}^2_{\gamma}/2M_{atom}~~,$$
and the gap energy, $\omega_{gap}$.
A rough estimate for $P_0$ is as follows:
\beq\label{bauer}
P_0~\sim \exp(-R/\omega_{gap})~.
\eeq
Thus, when the recoil energy is much larger than the gap, $R \gg \omega_{gap}$
$P_0$ tends to zero and kinematics is the same as for a free-atom
decay. In this limit we would come back to the parton-like picture.

The lattice data on the monopoles indicate that it is the opposite limit,
$R\ll \omega_{gap}$,
which can be relevant to the monopoles (if the analogy is correct at all).

\subsection{Quantum propagator as Brownian motion}

The M\"o\ss bauer effect concerns kinematics in the Minkowski space.
On the other hand, we are considering the monopole motions in the Euclidean space.
Moreover, there are no decays of the monopoles of course and,
at first sight, there is no physical quantity analogous
to ${\bf p}_{\gamma}$. However, we will argue in this subsection that
the famous Brownian picture~\cite{percolation}
for the quantum propagator provides a key
to identify an analog to $|{\bf p}_{\gamma}|$ in the monopole case.

The classical Brownian motion (see, {\it e.g.}, Ref.~\cite{percolation})
is the motion of a particle of mass $M_{Br}$
which results from exchange of momenta with particles of a
medium which are not observed and chaotic. Then the Brownian motion is a random
walk with the step
$$b~\sim~{\delta p \cdot\delta \tau\over M_{br}}~,$$
where $\delta \tau$ is the average time between collisions and $\delta p$
is the average momentum transfer during the collision.

In case of the quantum propagator one should rather think in terms
of the Brownian motion of a particle attached to a string with a
non-vanishing
tension. This is the origin of the inertia, or of the mass term
(for details see Sect. 4.1).
As for a non-vanishing $\Delta p$ it is now
quantum in origin and can be estimated from the uncertainty principle.
Since the monopole trajectory is measured on the scale $a$, generically
$\Delta p\sim a^{-1}$ and is parametrically large if $a\to 0$.

This simplest estimate should be corrected, however, for the fact that
one can introduce mass and kinetic energy only after coarse-graining
see Sect. 4.1. Therefore,
\beq\label{n}
\Delta p~\sim {1\over a\cdot\sqrt{n}}
\eeq
where $n$ was introduced in Eq. (\ref{constraint}). Note that $n$ does not depend
on $a$ and for practical estimates one can use
$$\sqrt{n}~\approx (2\div 3)~.$$

\subsection{Limit $a\to 0$: academic and realistic}

Coming back to our analogy with the M\"o\ss bauer effect, we conclude
that in the limit $a\to 0$ we expect that the existence of
the gap is not important and we can think in terms of the parton model.
Indeed, in the relativistic kinematics, $R\sim \Delta p\sim a^{-1}$
while the mass of ``free monopole'' $m_{prop}\sim a^{-1/2}$ and
can be neglected compared to $\Delta p$ at small $a$.
Moreover, the gap is provided by the glueball mass, $m_{gl}$ and does not
depend on
$a$ at all.

This conclusion could be foreseen: at short distances, i.e. in the limit
$a\to 0$ one can neglect the effects of the binding of the monopoles.
What is more surprising is that for  realistic
values of $a$ we are in fact far from
the academic limit $a\to 0$. This is due to interplay of various numerical
factors. Consider, for example, $a=(3 \GeV)^{-1}$ which corresponds to smallest
$a$ available. Then:
\beq\label{realistic}
m^2_0~\sim ~{8\mu\over a}~\sim~3~\GeV^2,~~(\Delta p)^2~\sim~2 \GeV^2
\eeq
where we used $n\approx 5$ (see Eq. (\ref{n})) and the chemical
potential~(\ref{mu}) can be estimated with the help of Eq.~(\ref{travel})
as $\mu\sim (1.6~fm)^{-1}$. Thus the ordering of $m_{prop}$
and $\Delta p$ is still reversed at the presently available lattices
compared to the expectations in the $a\to 0$ limit.

The onset of the parton model is expected when copious excitation  of
glueballs is favored. Since the glueball is relatively heavy,
$m_{gl}\sim 1.5 \GeV$ the realistic values of $\Delta p$
are far from satisfying the condition
\beq
(\Delta p)^2~\gg~ (m_{prop}+m_{gl})^2~.
 \eeq

Thus, we come to the following conclusion:

{\it The recoil-free geometrical picture of the
percolating  cluster is perfectly
consistent with the quantum mechanics
as far as the resolution (which is of order  $\sqrt{n}a$) is not too high.
At better resolution we expect to loose
both point-like monopoles and the geometrical
picture itself.}

Actually, in the data of Ref.~\cite{bornyakov01} one
can already see some hints on violations of the $1/a$ behaviour
of the monopole
field-theoretical mass $M(a)$.
Thus, we are inclined to correlate these first indications
to appearance of the monopole structure~\cite{bornyakov01}
and the scaling violations
in $d_{av}$, see Eq. (\ref{d}). Our prediction is that these
indications become clearer at smaller $a$.

One can roughly estimate the values of $a_{crit}$ at which the parton
picture begins to prevail over the recoil-less kinematics as:
\beq\label{thirty}
a^{-1}_{crit}~\sim~(10\div 30) \mbox{GeV}~.
\eeq
At such $a$ one can also expect that the monopoles do not look
point-like. Indeed, in the field theoretical language it is quite trivial
that the dominance of the inelastic processes and
appearance of the form-factor for an ``elastic'' process are
determined by the same dynamics.

Appearance of such a large mass scale as (\ref{thirty})
might come as a surprise.
It is worth noticing therefore that there exists accumulating, although
indirect evidence on the relevance of large mass scales
to the QCD phenomenology (see, {\it e.g.}, Ref.~\cite {massscale} and references therein).
Now we see how such a scale can be build up on $\Lambda_{QCD}$,
proceeding through the
glueball mass and other numerical factors.

\section{Monopoles and glueballs}
\label{six}

Thus, the next crucial question seems to be the coupling of the monopoles
to glueballs. In the most general way the monopole-glueball interaction can be studied
by measuring the correlation function:
\beq\label{correlator}
K(x-y)~=~\langle 0|\rho(x),\rho(y)|0\rangle~,
\eeq
where
$$\rho(t)\equiv \phi^{+}({\bf x},t)\phi({\bf x},t)~.$$
Asymptotic  at large distances is sensitive to a massive excitation~\cite{ivanenko}:
\beq\label{asymptotic}
\lim_{t\to\infty}{\langle 0|\rho(t),\rho(0)|0\rangle}~=~c_1~+c_2\,e^{-m_{gl}\cdot
t}\,,
\eeq
which is likely to be a glueball.

As far as we approximate monopoles as point-like {\it and} confine ourselves
to an effective $\lambda\phi^4$ theory {\it without gluons} then correlators like
(\ref{correlator}) are given by a sum over closed monopole loops:
\beq
K(x-y)~=~{\partial^2\over \partial j_1\partial j_2}Z(M)~{\Bigl|}_{j_1=j_2=0}
\eeq
where
$$  M(z)~=~ j_1\delta^4(x-z)+j_2\delta^4(y-z)~,$$
and the partition function is given by:
\beq
Z(M)~=~{1\over 4\pi^2}\int d^4x_0\int_0^{\infty}{dL\over L^3}e^{-m^2La}
\langle e^{-V(x(\tau))}\exp\big(-\int M(x(\tau))d\tau\big)\rangle_{x_0},
\eeq
where $\langle ...\rangle_{x_0}$ means averaging over all the closed paths.
Further details and definitions can be found, {\it e.g.}, in the review~\cite{schubert}.
What is important for us now is that the sum over the paths can be calculated numerically
using  the lattice data since the monopole trajectories are
directly observed.

The constant in Eq.~(\ref{asymptotic}) is related to the $\rho_{mon}$:
\beq
c_1~=~(8\rho_{mon}\cdot a)^2\,,
\eeq
and is entirely determined by the percolating cluster.
Determination of the glueball mass would require, on the other
hand, sensitivity to the quantum corrections, or
to the finite-size clusters\footnote{The percolating cluster alone at distances
$d\gg d_{av}$ is representing a cluster of fractal dimension $D_{fr}=4$
(see our discussion of the uncorrelated percolation in Sect.~\ref{two})
and  cannot be sensitive to the glueball mass.}.

It could quite well be so that keeping the closed paths alone would
not be adequate to determine the glueball mass through (\ref{asymptotic}).
This would be a signal that gluonic intermediate states are important
and the theory is not unitary without inclusion of such states.
Since the gluons are not
detected directly on the lattice the pure gluonic intermediate
states would look as breaks in the monopole loops.

Thus, measurements of the correlator (\ref{correlator}) would be very important to
further understand the structure of the effective theory of the monopole
interactions.

\section{Conclusions}
\label{seven}

We started with the picture according to which non-Abelian
gluodynamics, when
projected onto the scalar-field theory via monopoles, corresponds to a
fine tuned theory \cite{vz}.
The monopoles which we considered are defined (``detected'')
through the Maximal Abelian projection. However, the mass scales which
exhibit mass hierarchy are gauge independent. The scales are provided by
the $SU(2)$ invariant action per unit length of the monopole trajectory,
on one hand, and by inverse ``free path length'' (see (\ref{limits})), on the
other. The observed fine tuning suggests the use of the geometrical
language to describe the monopoles.

In this paper we confronted the polymer approach
to the effective theories of the magnetically charged fields
with the lattice data on
the monopole clusters.

There are two types of clusters, namely, finite-size and
percolating. In the former case we have demonstrated
that the length spectrum and the dependence of the size
of the clusters on their length are well understood
within the percolation theory. An alternative language is the simplest vacuum
loop in the polymer language.

There is a striking difference, however,
between the finite-size clusters and
the percolating cluster. In the former case one thinks in terms of
the vanishing mass of the monopoles (on the scale of the lattice spacing $a$):
\beq\label{zero}
(m^2_{mon})_{finite-size}~\approx~0\cdot a^{-2}.
\eeq
This picture is confirmed by evaluation of the critical exponents
just mentioned.
On the other hand, the straightforward interpretation of the
geometry of the percolating cluster leads to~\cite{boyko}:
\beq\label{infty} (m^2_{mon})_{percolating}~\sim~a^{-2}\,. \eeq

We
have argued that the apparent discrepancy between
the effective monopole masses extracted in the two ways
is naturally resolved if
one takes into account that the finite-size clusters correspond to
quantum corrections while the percolating cluster corresponds to
the ground state. The quantum corrections reveal then presence of
a massless (on the scale of $a$) excitation. In case of the ground
state, we observe an analog of the M\"o\ss bauer effect.
The onset of a parton-like description is delayed numerically
by some factors, most notably by a relatively high glueball mass $m_{gl}$
where $m_{gl}$ plays the role of the gap.

We also predict that at much smaller $a$ (see (\ref{thirty})) one
should see the convergence of (\ref{infty}) to (\ref{zero}).

Thus, first applications of the
polymer picture to the monopole clusters turn successful.
However, justification for such an approach
remains at present pure phenomenological in nature
and relies on the observation of the fine tuning.
Any further theoretical development is therefore to be checked by
measurements on the lattice.
In particular,
studies at smaller $a$, both theoretical and numerical,
are desirable.

\acknowledgments

We are grateful to D. Antonov, P.Yu. Boyko, F.V. Gubarev, D.
Diakonov, R. Hofmann, M. Koma, Y. Koma, K. Konishi, K. Langfeld,
G. Marchesini, S. Narison, M.I. Polikarpov, L. Stodolsky, T.
Suzuki, N. Uraltsev, and P. van Baal for discussions. M.N.Ch. is
supported by the JSPS Grant No. P01023. V.I.Z. is partially
supported ny the grant INTAS-00-00111 and by the DFG program
``From lattice to hadron phenomenology''.


\begin{thebibliography}{99}

\bibitem{classics}
Y. Nambu, {\it Phys. Rev.} {\bf D10}, 4262 (1974);\\
G. 't Hooft, {\it in} ``High Energy Physics'', Editorici Compositori, Bologna, (1975);\\
S. Mandelstam,  {\it Phys. Rep.} {\bf C23}, 516 (1976).

\bibitem{reviews}
M.N. Chernodub, M.I. Polikarpov,
{in "Cambridge 1997, Confinement, duality, and nonperturbative aspects of QCD"}, p.
387; hep-th/9710205;\\
T. Suzuki, {\it Prog. Theor. Phys. Suppl.} {\bf 131}, 633 (1998);\\
A. Di Giacomo, {\it Prog. Theor. Phys. Suppl.} {\bf 131}, 161 (1998).

\bibitem{action}
T. Suzuki, H. Shiba, {\it Phys. Lett.}, {\bf B 351}, 519  (1995);\\
S. Kato {\it et al}, {\it Nucl. Phys.} {\bf B 520}, 323 (1998);\\
M.N. Chernodub {\it et al}, {\it Phys. Rev.} {\bf D 62}, 094506 (2000).

\bibitem{alive}
M.N. Chernodub, F.V. Gubarev, M.I. Polikarpov , V.I. Zakharov,
 {\it Yad. Fiz.} {\bf 64}, 615 (2001), hep-th/0007135.

\bibitem{Bornyakov02} V.G. Bornyakov, M. Muller-Preussker, {\it Nucl. Phys. Proc.
Suppl.}, {\bf 106}, 646 (2002).


\bibitem{pochinsky}
T.L. Ivanenko, A.V. Pochinsky, M.I. Polikarpov,
{\it Phys. Lett.} {\bf B 252}, 631 (1990);\\
S. Kitahara, Y. Matsubara, T. Suzuki, {\it Progr. Theor. Phys.}
{\bf 93}, 1 (1995).

\bibitem{hart}
A. Hart, M. Teper, {\it Phys. Rev.}, {\bf B58}, 014504 (1998);\\
``Monopole spectra in non-ABelian gauge theories'',
hep-lat/9606022 (1996).

\bibitem{bornyakov01}
V.G. Bornyakov, et.al., {\it Phys. Lett.} {\bf B537}, 291 (2002);\\
V.A. Belavin, M.I. Polikarpov, A.I. Veselov, {\it JETP Lett.}
{\bf 74}, 453 (2001).

\bibitem{vz}
V.I. Zakharov, ``Hidden Mass Hierarchy in QCD'' , hep-ph/0202040 (2002).

\bibitem{parisi}
K. Symanzik, Proc. Int. School of Physics ``Enrico Fermi'',
Varenna Course XLV, ed. R. Jost, Academis Press (1969);\\
C.~A.~De Carvalhom, S.~Caracciolo, J.~Frohlich,  {\it Nucl.\ Phys.}
\ B {\bf  215}, 209 (1983);\\
G. Parisi, Statistical Field Theory, Addison-Wesley, (1987), Chapter 16.

\bibitem{percolation}
C. Itzykson, J.-M. Drouffe, {\it ``Statistical Field Theory''},
vol. 1, Cambridge University Press, (1989);\\
D. Stauffer, A. Aharony,
  {\it  ``Introduction to percolation theory''}
   London et al.: Taylor and Francis 1994;\\
G. Grimmett,
  {\it  ``Percolation''}
   Berlin et al.: Springer 1999,
   Grundlehren der mathematischen Wissenschaften, Vol. 321.


\bibitem{Polyakov75}
A.M. Polyakov, {\it Phys. Lett.}, {\bf B59}, 82 (1975).

\bibitem{BanksMyersonKogut}
T.~Banks, R.~Myerson and J.~B.~Kogut,
Nucl.\ Phys.\ B {\bf 129}, 493 (1977).


\bibitem{shiba}
H. Shiba, T. Suzuki, {\it Phys. Lett.}, {\bf B343}, 315 (1995).

\bibitem{samuel79} S. Samuel, {\it Nucl. Phys.}, {\bf B154}, 62 (1979).

\bibitem{ambjorn}
J. Ambjorn, B. Durhuus, T. Jonsson,
``Quantum Geometry. A Statistical Field Theory Approach'',
Cambridge, UK: Univ. Pr. (1997);\\
J. Ambjorn, ``Quantization of Geometry'',
talk given at Les Houches Summer School,  Les Houches, France,,
hep-th/9411179 (1994).

\bibitem{aizenman}
M. Aizenman, H. Kesten, C.M. Newman, {\it Comm. Math. Phys.} {\bf
111}, 505 (1987).

\bibitem{u1}
J. Jersak, C.B. Lang, T. Neuhaus, {\it Phys. Rev. Lett.} {\bf
D54}, 6909  (1996).

\bibitem{boyko}
P. Yu. Boyko, M.I. Polikarpov, V.I. Zakharov,
{\it ``Geometry of percolating monopole
clusters''}, hep-lat/0209075 (2002).

\bibitem{sch}
A. M. J. Schakel, {\it Phys. Rev.} {\bf E63} 026115, (2001);\\
S. Bund, A.M. J. Schakel, {\it Mod. Phys. Lett.} {\bf B13}, 349 (1999).


\bibitem{sasha}
A.M. Polyakov, ``Gauge Fields and Strings'',
Harwood Academic Publishes, Ch. 9 (1987).

\bibitem{stone78}
M. Stone,  P.R. Thomas, {\it Phys. Rev. Lett.}, {\bf 41} 351 (1978).

\bibitem{lipkin}
H.J. Lipkin, {\it Ann. Phys.}, (USA), {\bf 9}, 332 (1960);
{\bf 18} 182 (1962).

\bibitem{ivanenko}
T.L. Ivanenko, A.V. Pochinsky, M.I. Polikarpov, {\it Phys. Lett.}
{\bf B302}, 458 (1993);\\
K. Langfeld, H. Reinhardt, ``
Monopole-Anti-monopole in MAG projected lattice gauge theory'',
hep-lat/0206021 (2002).

\bibitem{massscale}
V.A. Novikov, M.A. Shifman, A.I. Vainshtein, V.I. Zakharov,
{\it Nucl. Phys.} {\bf B191} 301 (1981);\\
F.V.~Gubarev, M.I.~Polikarpov, V.I.~Zakharov,
{\it Phys. Lett.} {\bf B438} 147 (1998);\\
K.G. Chetyrkin, S. Narison, V.I. Zakharov,
{\it  Nucl. Phys.} {\bf B550} 353 (1999).

\bibitem{schubert}
Ch. Schubert, {\it Phys. Rept.} {\bf 355}, 73-234 (2001).

\end{thebibliography}
\end{document}